# BrightVAE: Luminosity Enhancement in Underexposed Endoscopic Images

Farzaneh Koohestani*, Zahra Nabizadeh*†, Nader Karimi*, Shahram Shirani†, Shadrokh Samavi†‡

*Department of Electrical and Computer Engineering, Isfahan University of Technology, Isfahan, Iran
†Department of Electrical and Computer Engineering, McMaster University, Hamilton, Canada
‡Department of Computer Science, Seattle University, Seattle, USA

**Abstract:** The enhancement of image luminosity is especially critical in endoscopic images. Underexposed endoscopic images often suffer from reduced contrast and uneven brightness, significantly impacting diagnostic accuracy and treatment planning. Internal body imaging is challenging due to uneven lighting and shadowy regions. Enhancing such images is essential since precise image interpretation is crucial for patient outcomes. In this paper, we introduce BrightVAE, an architecture based on the hierarchical Vector Quantized Variational Autoencoder (hierarchical VQ-VAE) tailored explicitly for enhancing luminosity in low-light endoscopic images. Our architecture is meticulously designed to tackle the unique challenges inherent in endoscopic imaging, such as significant variations in illumination and obscured details due to poor lighting conditions. The proposed model emphasizes advanced feature extraction from three distinct viewpoints—incorporating various receptive fields, skip connections, and feature attentions to enhance image quality and robustly support more accurate medical diagnoses. Through rigorous experimental analysis, we demonstrate the effectiveness of these techniques in enhancing low-light endoscopic images. To evaluate the performance of our architecture, we employ three widely recognized metrics—SSIM, PSNR, and LPIPS—specifically on the Endo4IE dataset, which consists of endoscopic images. We evaluated our method using the Endo4IE dataset, which consists exclusively of endoscopic images and showed significant advancements over the state-of-the-art techniques for enhancing luminosity in endoscopic imaging.

## 1- Introduction

The enhancement of image luminosity is particularly vital in the medical field, where it significantly contributes to more precise diagnostics by revealing subtle pathological markers often missed in underexposed images. This is especially important in endoscopic procedures, where the internal shape of organs can create dark areas that obscure crucial anatomical or pathological information [1]. As shown in Fig. 1, low-light conditions can limit the visibility of critical details, impacting the accuracy of diagnosis and formulation of treatment strategies. The variable lighting conditions typical in endoscopy present unique challenges, necessitating advanced luminosity enhancement techniques to ensure that medical professionals can accurately assess and respond to patient conditions [2], [3]. Thus, enhancing luminosity in endoscopic images is indispensable for improving diagnostic accuracy and facilitating better patient outcomes in the healthcare industry.

The images' clarity and expressiveness are paramount for tasks involving endoscopic medical image analysis. Endoscopic images captured under suboptimal lighting conditions often exhibit a range of issues, such as reduced contrast, uneven brightness levels, pervasive darkness in certain areas, and potential noise and blurring [4]. These challenges, illustrated in Fig. 1, can significantly degrade the utility and interpretability of these images, making the initial enhancement of image quality a crucial first step. Improving the quality of endoscopic images involves addressing these specific challenges to restore or enhance visibility and detail. This foundational improvement ensures that subsequent analyses or diagnostics are based on the most accurate and informative visual data possible [5], [6]. Clear and expressive images are essential in medical tasks aimed at diagnosing or assessing the state of diseases. This is especially true for endoscopic images and similar medical imaging modalities, where internal body visuals are often marred by uneven lighting and shadowy regions due to the inherently challenging conditions of capturing images within the body [1], [4].

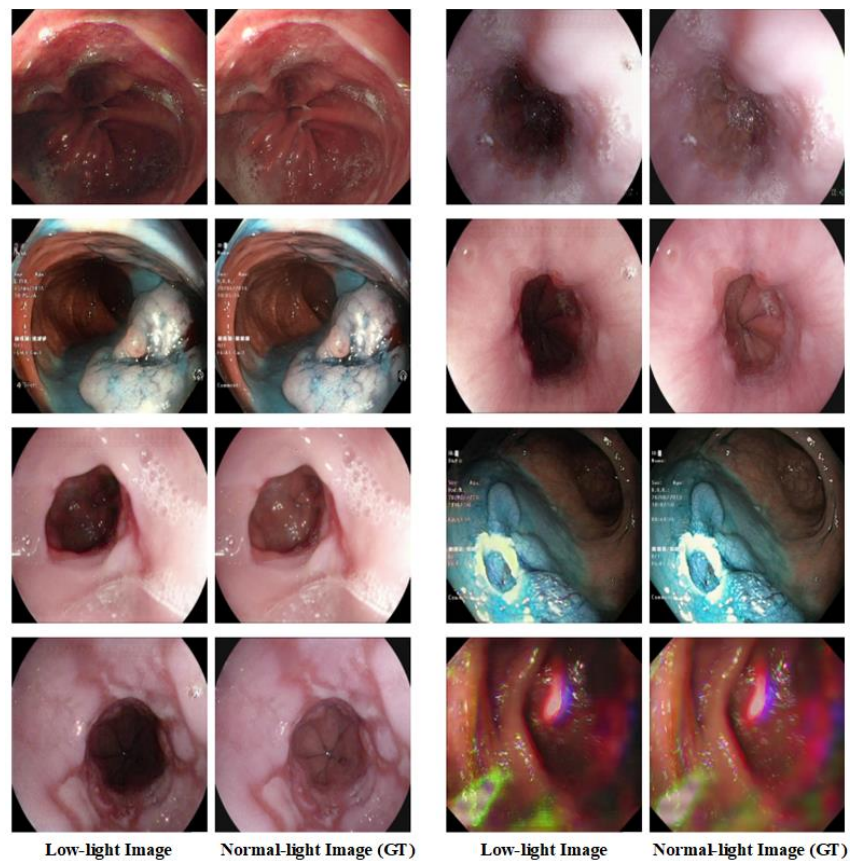

Fig. 1. Endoscopic Image Comparison: Low-light endoscopic images with corresponding normal-light (ground truth) images, highlighting challenges like reduced visibility and contrast loss. Ground truth images illustrate the need for luminosity enhancement to restore clarity and diagnostic accuracy.

Many general image enhancement methods fail with low-light images, such as those from endoscopic procedures. These methods may enhance overall brightness but struggle to reveal details in darker regions or could over-brighten lighter areas, leading to a loss of critical information and reduced contrast. As a result, these techniques are inadequate for images requiring nuanced visibility. This underscores the need for specialized methods to address the specific challenges of varying illumination in low-light endoscopic imaging.

Under various lighting conditions, endoscopic images often suffer from uneven lighting, reduced contrast, and pervasive shadowy regions that obscure detailed anatomical structures. The prevalence of such issues necessitates a preliminary enhancement of endoscopic image quality to ensure that visuals are sufficiently clear for accurate analysis. Once optimized, these images become invaluable tools in diagnosing critical conditions, such as various forms of cancer. The stakes in these medical contexts are profoundly high, as the clarity and accuracy of endoscopic images can directly influence diagnostic precision, treatment decisions, and, ultimately, patient outcomes. In scenarios where early detection and treatment can significantly alter the course of a disease, the role of enhanced medical imaging transcends technical necessity, becoming a pivotal factor that can influence life-or-death outcomes [5]. The proposed BrightVAE method utilizes a two-receptive field encoder-decoder structure combined with an advanced attention mechanism to create a new method of enhancing the luminosity of endoscopic images, ensuring clearer and more detailed visualizations.

This study's contributions are delineated across three distinct areas:

• The Attencoder module leverages an advanced attention mechanism to selectively amplify crucial features in low-light endoscopic images, enhancing visibility and detail by dynamically focusing on the most relevant regions.

• The Attenquant module revolutionizes vector quantization by integrating attention mechanisms that prioritize and preserve essential details, significantly improving data representation and image reconstruction accuracy.

• BrightVAE uses skip connections between the Attencoder and decoder to optimize feature transfer, enhancing clarity and detail in low-light medical imaging through a hierarchical architecture with dual-receptive fields.

This paper is structured as follows: Section II offers an overview of previous work in low-light image enhancement. Section III elaborates on our proposed method, BrightVAE. Section IV describes the experimental setup and discusses the results obtained from our experiments. Section V provides a conclusion and summarizes the findings of this research.

## 2- Related Works

In this section, we study related works, encompassing both traditional methods and those based on deep learning, to address the challenges of low-light image enhancement.

A. Traditional Methods

The advancement of low-light image enhancement techniques is categorized into histogram-based and retinex-based methods. In histogram-based techniques, innovations such as dynamic histogram equalization [7] optimize image contrast by independently adjusting segmented histograms. In contrast, dualistic sub-image histogram equalization [8] splits the image based on median brightness for tailored enhancements. Recursive mean-separate histogram equalization [9] refines these enhancements by maintaining brightness balance. Retinex-based methods, drawing from human visual perception under varying lighting, include automated multiscale retinex [10], which adjusts enhancement dynamically across image segments, and the robust retinex model [11] that integrates noise variance estimation to improve detail preservation in low-light images.

B. Deep Learning Methods

This section explores deep learning methods applied to enhance image quality, focusing on both medical and general applications. We will examine techniques specifically developed for medical imaging, such as endoscopic image enhancement, alongside general methods that address low-light conditions in various settings. These advanced techniques are tailored to improve visibility and detail in low-light environments, specifically catering to the unique requirements of medical diagnostics and broader imaging challenges.

1) Medical Methods: Several contributions have been made to improve image quality under poor lighting and exposure conditions in medical imaging enhancement. Bai et al. [6] introduce EndoUIC, a promptable diffusion transformer capable of adaptively correcting illumination in capsule endoscopy, significantly improving image quality across varying exposure levels.

García-Vega et al. [1] have refined the LMSPEC method, extending its application to correct over- and underexposure while preserving structural details in endoscopic images. Similarly, Mou et al. [5] developed a process that enhanced global illumination and local information through a module with dual attention mechanisms, specifically for low-light endoscopic images. Yue et al. [4] propose DPENet, utilizing a multiscale feature extraction strategy to boost visibility under poor lighting conditions. Lastly, the introduction of StillGAN [12] employs a bi-directional GAN to enhance medical image quality by applying structural and illumination constraints, effectively distinguishing between low- and high-quality image domains.

These developments represent a significant advancement in medical image enhancement, supporting more accurate clinical analysis and diagnosis.

2) General Methods: Recent advancements in image enhancement technology have effectively combined sophisticated AI techniques with traditional methods, achieving significant improvements in image quality. LLFlow method by Wang et al. [13] uses normalizing flows to model light distribution effectively. Cui et al. [14] introduced the Innovative Adaptive Transformer, focusing on adaptive lighting adjustments and marking a shift towards deeper AI integration. The CSLIE method enhances low-light images without a reference, making it adaptable to varied lighting conditions [15]. Zhang et al.'s Knid++ method also preserves the natural appearance in low-light enhancements [16]. At the same time, Liu et al. [17] developed an iterative network that alternates between enhancement and fusion steps to restore and clarify images. The FRLIE network [18] employs an encoder-decoder structure with a feature restoration subnet to transform low-light image features into clear images, effectively reducing noise and color distortion. FLW [19] enhances low-light images through global feature extraction and design optimizations, addressing brightness, contrast, and color consistency. The LMSPEC method [20] utilizes a coarse-to-fine deep neural network to correct exposure errors, improving over-exposed and under-exposed images.

MBPNet [21] applies a multi-branch architecture with progressive enhancement using convolutional LSTMs and a multi-loss function for low-light image enhancement across multiple scales. Finally, VQLIE [22] employs a VQ-VAE architecture to enhance low-light images, adding another layer of sophistication.

These diverse approaches underline a dynamic evolution of learning-based methods in image processing. By leveraging deep learning, these methods not only enhance the capabilities of traditional image processing techniques but also open new avenues for efficient and effective image enhancement and restoration across various domains. Despite these advancements, existing methods often struggle with maintaining detail preservation while enhancing illumination uniformly. BrightVAE addresses these gaps by integrating multiscale feature attention mechanisms, offering a balanced approach to global and local enhancements.

## 3- Proposed Method

In this research, we present BrightVAE, a novel approach inspired by the hierarchical Vector Quantized Variational Autoencoder (VQ-VAE) architecture [23], tailored specifically for enhancing low-light endoscopic images. At the core of BrightVAE's innovation lies the Attencoder and Attenquant modules, which incorporate sophisticated attention mechanisms to enhance the processing of features significantly. The Attencoder module targets and amplifies critical regions within images, optimizing feature extraction precisely where it is most needed. Concurrently, the Attenquant module leverages attention-driven quantization to preserve vital details through the processes of image compression and reconstruction, ensuring high perceptual quality.

Key to our approach is the integration of the Attencoder and Attenquant modules, which leverage dual-receptive fields to capture both broad, contextual features and intricate local details through an attention-driven process. In this dual-receptive setup, our method utilizes two branches—a global branch for overall anatomical structures and a local branch for fine details—working together to enhance low-light endoscopic images. Combined with well-designed skip connections between the Attencoder and Decoder, these modules ensure that essential features are effectively transferred and utilized in the output image. This approach enables BrightVAE to address the unique challenges of medical imaging, where both large anatomical structures and critical minute features must be clearly discernible. The dual-receptive field design ensures that our architecture can enhance overall image brightness and detail without sacrificing any essential information, making it particularly effective for low-light conditions. Figure 2 illustrates this architecture, with further explanations to be provided in the subsequent sections.

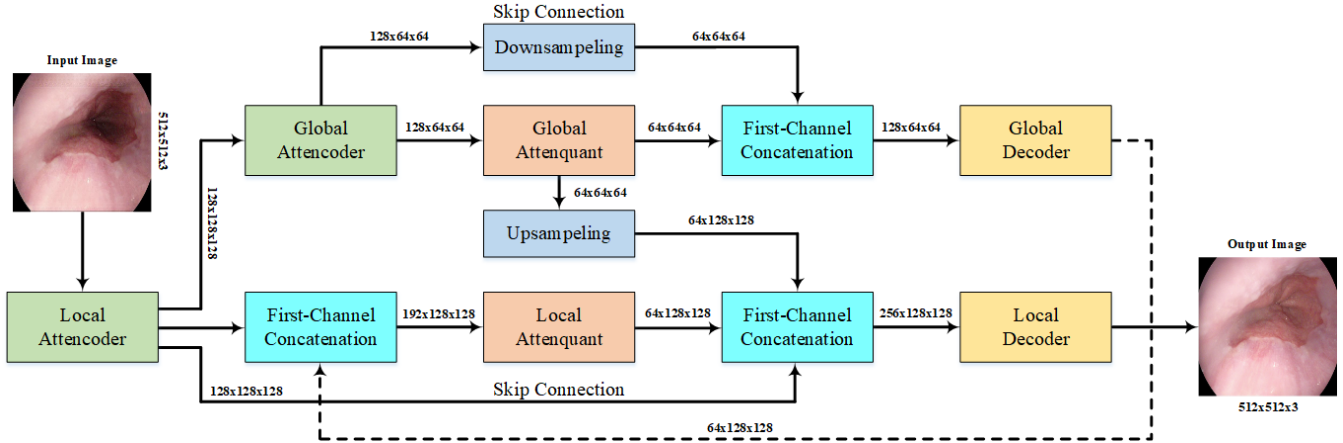

Fig. 2. Block diagram of BrightVAE, featuring dual processing paths. The top pathway displays the Global Attencoder, Attenquant, and Decoder, focusing on global image enhancements. The bottom pathway presents the Local Attencoder, Attenquant, and Decoder, specialized for local enhancements in endoscopic images.

A. Attencoder

The Attencoder module utilizes an attention mechanism to adaptively enhance low-light images by selectively emphasizing essential features. This attention-driven approach enables the module to dynamically allocate focus, prioritizing key areas for comprehensive scene understanding and detail refinement. By leveraging attention, the Attencoder can effectively enhance image clarity and reduce noise while preserving contextual integrity and fine details, essential for medical diagnostics. The structure of this module is shown in Fig. 3.

The Attencoder starts with the initial block, which is configured differently depending on the receptive field of operation. For the global branch, the initial block utilizes a streamlined arrangement of two 3x3 convolutional layers with a ReLU activation between them, focusing on broad, contextual features. In contrast, for local branch operations aimed at local branch, the initial block comprises three 3x3 convolutional layers, each followed by a ReLU activation, enhancing its capability to process and refine finer image details [23].

Following the initial block, the Attencoder includes two ConvResBlock modules, which subtly enhance feature maps by adding depth to the network while avoiding vanishing gradients through residual connections. Initially introduced in the VQ-VAE method [23], the ConvResBlock integrates seamlessly into BrightVAE, employing ReLU activations, convolutional layers, and skip connections to support effective feature processing without batch normalization. This streamlined architecture maintains robust training and convergence characteristics, as skip connections promote gradient flow, enabling efficient learning in deeper layers. These features contribute to the stability and performance of BrightVAE, which is crucial for low-light image enhancement, helping retain essential details across deeper layers and supporting both local detail recovery and global feature synthesis. A ReLU activation layer follows, providing non-linearity to the processing flow, which helps break up the operations' linearity and allows the network to learn more complex patterns within the image data.

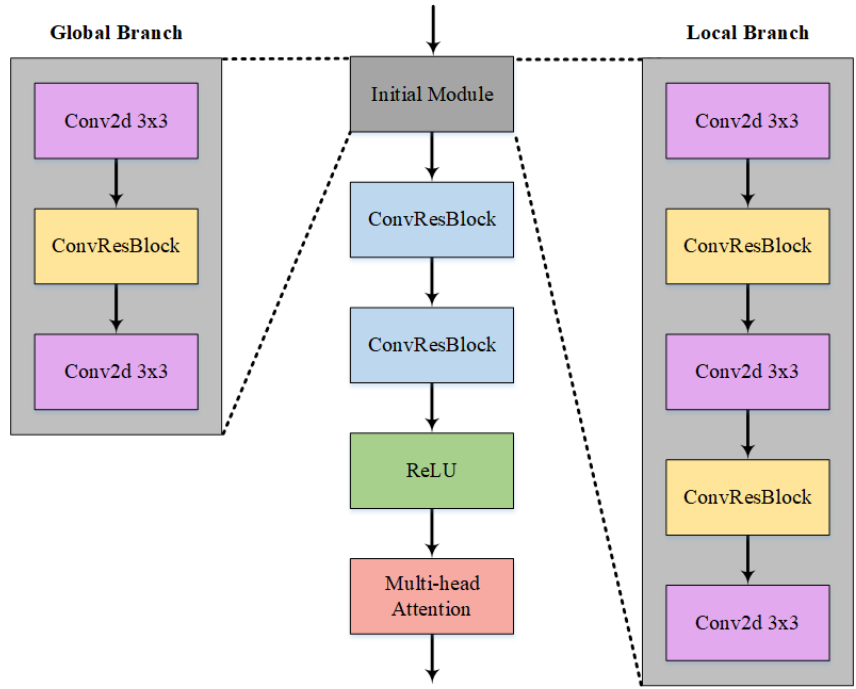

Fig. 3. Diagram of the proposed Attencoder architecture.

The final component in the Attencoder module is the Multi-head Attention mechanism. This layer is particularly effective in integrating information across various parts of the image. By focusing on how different regions of the image relate to each other, the attention mechanism can highlight features that are more relevant for enhancing the overall image quality. This is beneficial for both receptive fields as it helps fine tune the enhancements by aligning local details with global contextual insights.

When operating on a local receptive field, the Attencoder module generates detailed, normal-light reconstructions from low-light inputs, effectively enhancing visibility and clarity. The detailed configuration of the initial block ensures that finer textures and contrasts are captured and enhanced. Meanwhile, the global receptive field operation leverages a simplified initial block setup to prioritize broad feature contexts, which are essential for understanding the overall scene.

This dual-receptive functionality makes the Attencoder a robust solution for low-light image enhancement, capable of addressing both the micro-level details required for high-quality image restoration and the macro-level contexts needed for comprehensive scene understanding. By dynamically adjusting its internal structures and processing flows, the Attencoder ensures optimal handling of diverse image enhancement challenges, making it a versatile tool in advanced visual processing tasks.

B. Attenquant

The Attenquant module introduces an advanced approach to vector quantization by incorporating an attention mechanism into the quantization process, as illustrated in Fig. 4. This process starts with the Feature Map Input, which undergoes a series of linear transformations and LeakyReLU activations. These transformations, inspired by the original module from [23], project the input into a higher-dimensional space and then back down, enabling the model to capture complex dependencies within the data.

Building on this structure, we developed the Attention Score computation to assign significance to each feature in the vector, determining which features should be prioritized during quantization. These scores are then normalized using a softmax function in the attention weight computation step, ensuring the sum of the weights is one and distributing focus across input features based on their relative importance. This produces the Attention Weighted Feature Vector, which adapts the feature influence based on calculated importance, enhancing model precision.

Subsequent steps follow the quantization process outlined in [23]. The squared Euclidean distance is calculated between each input vector and embedding vectors, with attention weights influencing these distances to bias selection toward more crucial features. Optimal embeddings are then identified by the embedding index selection module, as illustrated in Fig. 4.

This involves a lookup in the distance matrix to approximate each input vector with the most suitable embedding from the dictionary. Finally, embeddings retrieved by the embedding retrieval module replace the original input vectors, producing the Attention Quantized Output. This output retains essential information while reducing dimensionality and complexity, enhancing the fidelity of quantized representations for greater adaptability across complex datasets.

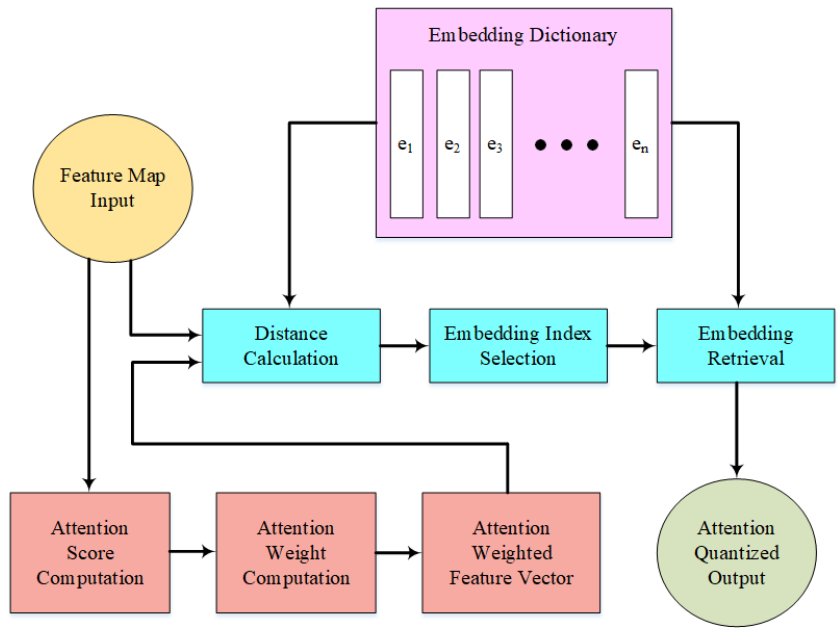

Fig. 4. Diagram of the proposed Attenquant architecture.

C. Decoder

The BrightVAE model's decoder module, introduced as part of the hierarchical VQ-VAE architecture [23], is expertly designed to perform dual functions depending on its receptive field: global branch and local branch. This flexibility is crucial for delivering high-quality outputs in the context of low-light image enhancement. The module begins with an initial convolutional layer that transitions from the input channel to a broader channel configuration, setting the stage for comprehensive processing tasks.

On one receptive field, the decoder is primarily tasked with global feature reconstruction [23]. It utilizes multiple instances of ConvResBlock blocks to refine and enhance the feature maps, capturing broad and essential features that represent the overall context of the image. This function is vital for understanding the global dynamics within the scene, which informs further processing stages.

In other receptive fields, the decoder focuses on local feature reconstruction. Here, the ConvResBlock blocks are crucial for ensuring that finer details and textures lost during earlier stages are effectively reconstructed, restoring image quality and sharpness. A ReLU activation follows these blocks to introduce non-linearity, enhancing the decoder's ability to model complex functions necessary for detailed endoscopic image reconstruction. The final part of the decoder employs transposed convolutional layers to upscale the processed features back to the original image dimensions.

D. Dual-Receptive Processing and Skip Connection

The model employs a two-receptive architectural approach that enhances its performance and effectiveness. The encoder processes the original image in this hierarchical setup on one receptive field. Then, the output is fed into a secondary network that operates at a different receptive field. This strategy is pivotal for capturing a comprehensive range of features by ensuring that features are optimally encoded and quantized at various resolutions, thus enhancing the overall quality and precision of the output.

Additionally, the model integrates the output from the second-receptive decoder with the quantized output from the first-receptive before recycling it back to the first-receptive decoder. This integration enhances the model's ability to reconstruct high-fidelity images. By compiling and refining details across different receptive fields, this method not only improves resolution and robustness but also accelerates convergence, boosting performance across various image-processing tasks. This makes the model highly adaptable and effective for tackling complex challenges.

Incorporating a skip connection between the encoder and decoder at each receptive field has significantly improved the model's performance, as demonstrated in Fig. 2. This architectural enhancement facilitates the direct transfer of feature maps from the encoder to the decoder, preserving essential spatial and contextual information that often diminishes during deep network processing. Skip connections prove invaluable in tasks requiring precise reconstruction of input data from its compressed form, such as image segmentation or detailed image restoration. By linking the encoding and decoding phases, skip connections not only accelerate the network's convergence but also enhance the accuracy and integrity of the outputs. This improvement leverages the strengths of both the encoder and decoder, optimally utilizing learned features across various levels of abstraction to boost the model's robustness.

E. Loss Function

In our comprehensive approach to endoscopic image enhancement, the total loss function incorporates three distinct components, each targeting specific aspects of image quality. Here is a detailed breakdown of each component, along with their formulas:

1) Image Restoration Loss (LRest): This loss function is crucial for ensuring that the reconstructed image closely matches the original in terms of pixel values. It is calculated using the Mean Squared Error (MSE), a common measure in image processing that quantifies the difference between the enhanced and original images:

$$L_{\text{Rest}} = \frac{1}{N} \sum_{i=1}^{N} (y_i - \hat{y}_i)^2$$

Here, $y_i$ represents the pixel values of the original image, $\hat{y}_i$ those of the reconstructed image, and $N$ is the total number of pixels. This loss emphasizes accurate pixel reproduction, which is fundamental to maintaining image fidelity.

2) Latent Space Optimization Loss ($L_{\text{Latent}}$): Following the approach used in [23], we apply a latent loss to ensure that the encoded features closely match the nearest vectors in the embedding space, promoting consistency and stability in the encoding process. The formula for this loss is:

$$L_{\text{Latent}} = \|\text{sg}[z_e(x)] - e\|^2 + \beta \|z_e(x) - \text{sg}[e]\|^2$$

where $Z_e(x)$ denotes the encoder output, e is the closest embedding vector, sg[·] denotes the stop-gradient operator to prevent gradients from flowing through specific paths, and β is a coefficient balancing the two terms. This loss helps stabilize the feature representation, making the network's output more reliable and consistent.

3) The third section of the loss function, $L_{\text{Similarity}}$, calculates the similarity between the output and ground truth images. Eight different loss functions were evaluated in subsection 4.D.2 of the ablation study to identify the appropriate loss function for assessing similarity.

Each loss component is weighted appropriately in the total loss function to ensure a balanced training process. For our proposed model, these coefficients are set as $\lambda_{\text{Rest}} = 1.0$, $\lambda_{\text{Latent}} = 0.25$, and $\lambda_{\text{Similarity}} = 0.08$:

$$L_{\text{Total}} = \lambda_{\text{Rest}} L_{\text{Rest}} + \lambda_{\text{Latent}} L_{\text{Latent}} + \lambda_{\text{Similarity}} L_{\text{Similarity}}$$

## 4- Experimental Results

In this section, we review the experimental results, including implementation details, dataset and metrics used, and comparisons with other methods. Additionally, an ablation study analyzes component impacts and the effects of different loss functions on model performance.

A. Implementation Details

In this study, the networks were developed using the PyTorch framework. The training spanned over 1000 epochs, utilizing the ADAM optimizer for optimization. A cyclic learning rate strategy was implemented, characterized by an initial ramp-up during the warm-up phase, followed by regular oscillations for the remainder of the training. Consistency was maintained in the image sizes throughout the experiments, and a batch size of 5 was used. It is important to note that all computational experiments were conducted on an NVIDIA GTX 3090 GPU, supported by a RYZEN 9 3900XT processor, to ensure substantial computational capability.

B. Dataset and Metrics

In this study, we utilized a publicly available dataset, Endo4IE [3], comprised of endoscopic images under low-light conditions. The dataset facilitates comparing high-quality, standard illumination images and their corresponding low-light synthetic counterparts. These synthetic images were generated utilizing a sophisticated pre-trained generative model.

The Endo4IE dataset is structured into two subsets: a training set containing 690 image pairs and a testing set comprising 266 images. Each image within the dataset has a resolution of $512 \times 512$ pixels, ensuring a consistent format for analysis and evaluation.

To evaluate the performance of our proposed method and compare it with other methods, we selected the Peak Signal-to-Noise Ratio (PSNR), the Structural Similarity Index Measure (SSIM) [24], and the Learned Perceptual Image Patch Similarity (LPIPS) [25] as our evaluation metrics. PSNR is employed to gauge the quality of reconstructed images through pixel intensity similarities. SSIM evaluates the impact on visual elements such as texture and structure. LPIPS, a metric based on deep learning, measures perceptual similarity, aligning closely with human visual perception. These metrics were chosen to analyze our method relative to competing approaches comprehensively.

C. Comparison with State-of-the-art

To compare our proposed model, BrightVAE, with existing techniques, we conducted a comprehensive evaluation using the Endo4IE dataset, a benchmark widely recognized in endoscopic low-light image enhancement. Our study involved an extensive comparison with a broad array of leading techniques, including KinD++ [16], EFINet [17], LLFlow [13], IAT [14], FRLIE [18], FLW [19], LMSPEC [20], VQLIE [22], MBPNet [21], StillGAN [12], EndoUIC [6], DPENet [4], Garcia's [1], and En Mou's [5]. Additionally, since our model draws inspiration from VQVAE [23], we include it in this comparison to underscore BrightVAE's enhancements to endoscopic low-light image enhancement. The results of this thorough evaluation are detailed in Table 1, clearly highlighting BrightVAE's advanced performance when benchmarked against these established methods. In Table 1, G means "General", and M refers to "Medical" images.

As demonstrated, our newly developed BrightVAE exhibits notable performance improvements across various metrics. Specifically, when compared to the recent En Mou's [5] method, BrightVAE outperforms it with an increase of about 3.252 dB in PSNR, an improvement of 0.045 in SSIM, and a reduction of approximately 0.014 in LPIPS on the Endo4IE Dataset. Additionally, when compared to the method from which we drew inspiration, VQVAE [23], BrightVAE shows a significant advantage, achieving 3.337 dB higher in PSNR, a 0.03 improvement in SSIM, and a 0.038 reduction in LPIPS. Notably, BrightVAE surpasses all other models in three key metrics—PSNR, SSIM, and LPIPS—for low-light image enhancement, securing top performance on the Endo4IE dataset. This demonstrates BrightVAE's superior capability to enhance low-light images compared to other methods. Figure 5 illustrates the effectiveness of our enhancement technique by displaying endoscopic images in original low-light conditions, after enhancement, and in normal lighting (GT) for comparison.

Table 1: Comparing results of various methods and those of ours.

| Method | Class | PSNR ↑ | SSIM ↑ | LPIPS ↓ |
|---|---|---|---|---|
| KinD++ [16] | G | 26.246 | 0.819 | 0.225 |
| EFINet [17] | G | 23.128 | 0.773 | 0.185 |
| LLFlow [13] | G | 25.924 | 0.794 | 0.236 |
| IAT [14] | G | 25.933 | 0.805 | 0.227 |
| FRLIE [18] | G | 27.590 | 0.818 | 0.243 |
| FLW [19] | G | 25.607 | 0.819 | 0.149 |
| LMSPEC [20] | G | 24.204 | 0.757 | - |
| VQLIE [22] | G | 26.364 | 0.843 | 0.178 |
| MBPNet [21] | G | 27.233 | 0.806 | 0.168 |
| VQVAE [23] | G | 27.239 | 0.849 | 0.171 |
| Garcia's [1] | M | 24.201 | 0.792 | - |
| StillGAN [12] | M | 25.571 | 0.776 | 0.206 |
| EndoUIC [6] | M | 25.490 | 0.852 | 0.194 |
| DPENet [4] | M | 26.803 | 0.849 | 0.213 |
| En Mou's [5] | M | 27.324 | 0.834 | 0.147 |
| **Ours** | **M** | **30.576** | **0.879** | **0.133** |

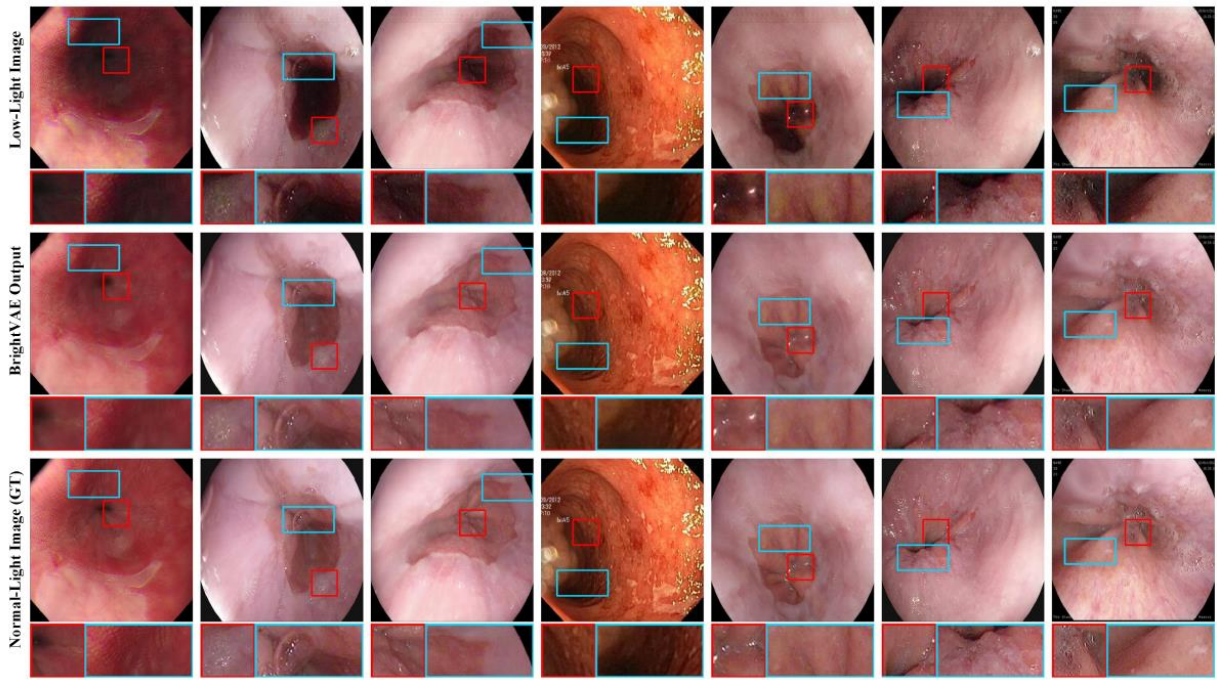

Fig. 5. Comparative display of endoscopic images of Endo4IE [3]: the first row shows original low-light images, the second row presents images enhanced by our method, and the third row contains reference images under normal lighting (Ground truth), demonstrating the effectiveness of our low-light enhancement technique.

D. Ablation Study

In this section, we conduct an ablation study to explore the influence of various components and loss functions within our model, assessing their individual contributions to overall performance. This focused analysis allows us to pinpoint enhancements and optimize our approach effectively.

1) Analysis of Components and Structure: This section examines the impact of modifying components in the baseline network. Replacing the baseline encoder with our Attencoder and the quantization module with our Attenquant module significantly improves model efficiency. Table II highlights the performance benefits of connecting the Attencoder directly to the final decoder, enhancing low-light processing in endoscopic images. Additionally, the processing of features at two receptive fields yields superior results over a single-receptive setup. Figure 6 visually compares models, displaying improvements in brightening dark areas of endoscopic images, confirming the effectiveness of our architectural enhancements. In Figure 6, the top left image is under low-light conditions, followed by outputs from six different models as detailed in Table 2, with the bottom right image displaying the same scene under normal lighting (the ground truth), illustrating the varying effectiveness of each model in enhancing visibility.

Table 2. Enhancement of model configuration illustration of improvements in PSNR, SSIM, and LPIPS metrics.

| Configuration | Two Receptive Fields | SSI Loss | Skip Connection | Attencoder Module | Attenquant Module | PSNR ↑ | SSIM ↑ | LPIPS ↓ |
|---|---|---|---|---|---|---|---|---|
| 1 | ✗ | ✗ | ✗ | ✗ | ✗ | 26.120 | 0.829 | 0.187 |
| 2 | ✓ | ✗ | ✗ | ✗ | ✗ | 26.167 | 0.842 | 0.190 |
| 3 | ✓ | ✓ | ✗ | ✗ | ✗ | 26.364 | 0.844 | 0.178 |
| 4 | ✓ | ✓ | ✓ | ✗ | ✗ | 27.313 | 0.842 | 0.188 |
| 5 | ✓ | ✓ | ✓ | ✓ | ✗ | 30.425 | 0.879 | **0.129** |
| 6 | ✓ | ✓ | ✓ | ✓ | ✓ | **30.576** | **0.879** | 0.133 |

2) Assessing the Impact of Various Loss Functions on Model Performance: In this section, we evaluate various loss functions to optimize our model's performance in determining the similarity between reconstructed images and ground truth (LSimilarity) for endoscopic low-light enhancement. This analysis identifies metrics that best capture image fidelity and accuracy under low-light conditions, helping us choose the most effective loss functions to ensure robust and reliable enhancements.

a) Jaccard Loss: The Jaccard Loss function [26], based on the Intersection over Union (IoU) metric, is used in image segmentation to evaluate model accuracy. It adjusts predictions to a scale of 0 to 1, simplifies them into one-dimensional arrays, and then measures the overlap between these predictions and the actual targets. The loss is calculated as $1 - IoU$, where a lower loss indicates better alignment between the predicted and actual areas. This loss helps refine model accuracy by emphasizing the maximization of overlap with true data regions.

b) Total Variation Loss: The Total Variation Loss (TV Loss) function [27] improves image quality by smoothing spatial variations while preserving edges. It minimizes squared differences between adjacent pixels, enhancing image smoothness and brightness without causing graininess or color shifts, thereby maintaining the image's natural look.

c) Cosine Similarity Loss: The Cosine Similarity Loss function [28] evaluates the alignment of feature vectors by converting the cosine of angles between vectors into a loss metric. Values close to one indicate better alignment. This function is effective for enhancing dimly lit images, ensuring that enhancements preserve original textures and minimize visual distortions, maintaining essential image characteristics.

d) Image Kullback-Leibler Divergence Loss: The Image Kullback-Leibler Divergence Loss function [29] enhances dimly lit images by minimizing the divergence between their pixel intensity distributions. It converts images into normalized histograms and uses Kullback-Leibler divergence to measure deviations. This process improves lighting while preserving critical features like contrast and sharpness, ensuring the natural appearance of the original image is maintained.

e) Gradient Magnitude Similarity Deviation Loss: The Gradient Magnitude Similarity Deviation (GMSD) Loss function [30] enhances image clarity by using Sobel filters to calculate gradient magnitudes in images, assessing these gradients through the Gradient Magnitude Similarity (GMS) metric. The GMSD loss, calculated by subtracting the similarity score from one, indicates how well the image preserves edge details and natural appearance during enhancement. Lower scores mean better preservation and visibility improvement.

f) Perceptual Loss: The Perceptual Loss function [25] uses a pre-trained VGG network to enhance image quality by matching high-level features like textures and colors to a target image. It compares feature maps from the VGG using MSE to minimize perceptual differences, effectively adjusting brightness and contrast to improve clarity and coherence while maintaining the original aesthetic appeal.

g) Color Consistency Loss: The Color Consistency Loss function [31] ensures color fidelity in image enhancements by aligning the mean and variance of color distributions between output and ground truth images. It minimizes mean squared differences to maintain consistent color tones and prevent distortions, thus preserving the natural look of the original scene while improving visibility.

h) SSI Loss: The Structural Similarity Index Loss function [24] measures structural and perceptual similarity between enhanced and original images, focusing on essential visual elements like texture, contrast, and structure. It evaluates changes in normalized pixel intensities, enhancing the understanding of image quality. The SSIM loss is calculated as:

$$L_{\text{SSI}} = 1 - \text{SSIM}(y, \hat{y})$$

The SSIM prioritizing features significant in human perception for image enhancement algorithms.

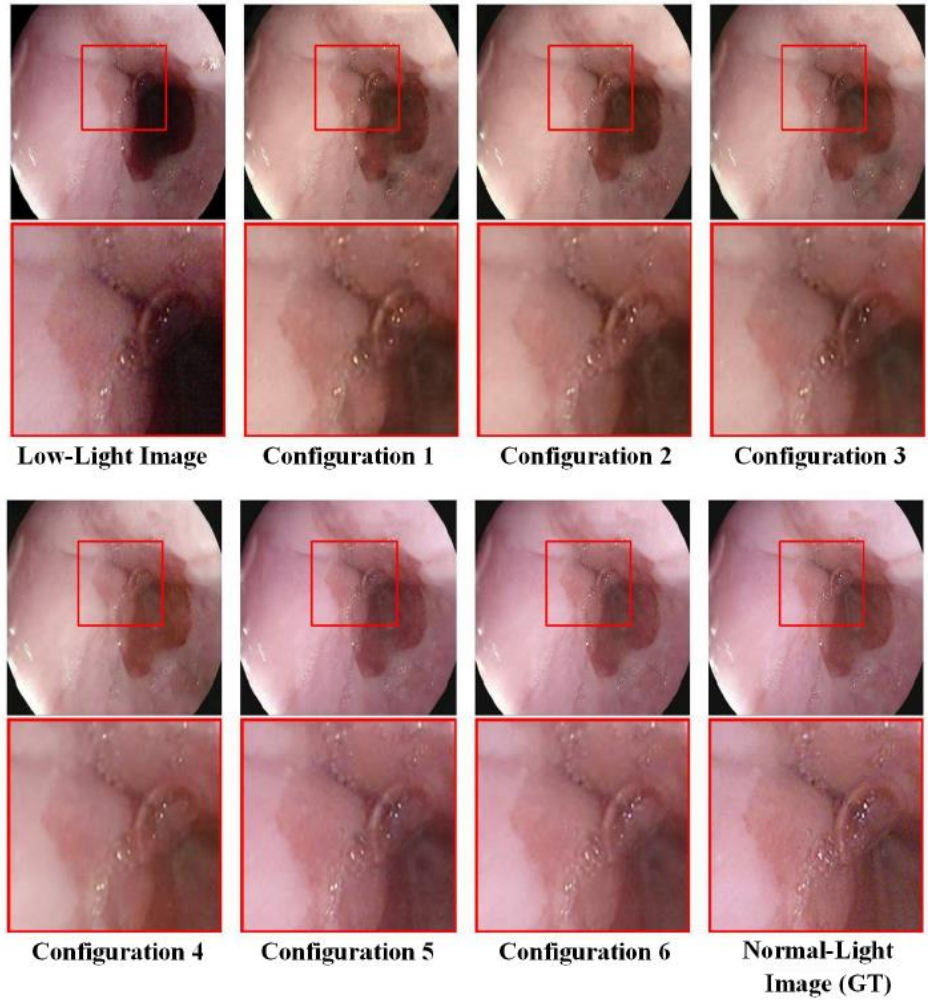

Fig. 6. Sequential comparison of endoscopic image enhancements.

Table 3. Comparison of metrics for various loss functions.

| Loss Function | PSNR ↑ | SSIM ↑ | LPIPS ↓ |
|---|---|---|---|
| Rest & Latent | 28.556 | 0.871 | 0.134 |
| Rest & Latent & Jaccard | 23.394 | 0.859 | 0.143 |
| Rest & Latent & TV | 25.391 | 0.731 | 0.4750 |
| Rest & Latent & Cosine Similarity | 27.564 | 0.846 | 0.170 |
| Rest & Latent & KLD | 27.935 | 0.864 | 0.158 |
| Rest & Latent & GMSD | 28.098 | 0.828 | 0.151 |
| Rest & Latent & Perceptual | 28.555 | 0.845 | **0.108** |
| Rest & Latent & Color Consistency | 28.799 | 0.857 | 0.167 |
| **Rest & Latent & Structural Similarity** | **30.576** | **0.879** | 0.133 |

Our analysis of various loss functions for enhancing dimly lit endoscopic images reveals that each function offers unique benefits. Still, the SSI loss function stands out as the most effective overall. Table 3 shows that SSI achieves the highest metric scores in SSIM at 0.879 and PSNR at 30.576, along with a strong LPIPS score of 0.133, which is the second best among the evaluated loss functions. These results highlight SSI's superior ability to preserve perceptual quality and visual fidelity, making it the preferred choice despite a slightly lower LPIPS score.

# 5- Conclusion

The contributions and methodology outlined in this study address a critical need for enhanced image quality in medical imaging, particularly in the challenging context of endoscopic and similar modalities. By unveiling BrightVAE and its innovative modules, namely the Attencoder and Attenquant, this research presents a comprehensive solution for luminosity enhancement, tackling issues of uneven lighting and shadowy regions prevalent in endoscopic images. The significance of this work lies in its dual impact: improving image brightness and enhancing diagnostic precision in medical contexts where image clarity directly affects patient outcomes. The introduction of BrightVAE marks a significant advancement in low-light image enhancement, thanks to its hierarchical approach that meticulously scrutinizes images at both local and global receptive fields. Moreover, the Attencoder and Attenquant modules show the efficacy of attention mechanisms in identifying and enhancing dark areas, while simultaneously quantizing features to improve precision in data representation. Each contribution to the new architecture is thoroughly analyzed, and experimental results validate their importance and effectiveness.